# Impact of Gd doping on morphology and superconductivity of NbN sputtered thin films

Rajveer Jha, Jeevan Jyoti and V. P. S. Awana

Quantum Phenomena and Applications Division, National Physical Laboratory (CSIR)

Dr. K. S. Krishnan Raod, New Delhi-110012, India

## Abstract

We report effect of Gd inclusion in the NbN superconductor thin films. The films are deposited on single crystalline Silicon (100) by DC reactive sputtering technique i.e., deposition of Nb and Gd in presence of reactive $N_2$ gas. The fabricated relatively thick films (400 *nm*) are crystallized in cubic structure. These films are characterized for their morphology, elemental analysis and roughness by Scanning Electron Microscopy (SEM), Energy Dispersive X-ray spectroscopy (EDAX) and Atomic Force Microscopy (AFM) respectively. The optimized film (maximum $T_c$) is achieved with gas ratio of Ar:$N_2$ (80:20) for both pristine and Gd doped films. The optimized NbN film possesses $T_c$ ($R=0$) in zero and 140kOe fields are at 14.8K and 8.8K respectively. The Gd doped NbN film showed $T_c$ (R=0) in zero and 130kOe fields at 11.2K and 6.8 K respectively. The upper critical field $Hc_2(0)$ of the studied superconducting films is calculated from the magneto-transport [$R(T)H$] measurements using GL equations. It is found that Gd doping deteriorated the superconducting performance of NbN.

*Corresponding Author
Dr. V. P. S. Awana, Senior Scientist
E-mail: awana@mail.npindia.org
Ph. +91-11-45609357, Fax-+91-11-45609310
Homepage www.fteewebs.com/vpsawana/

# Introduction

Although the cuprate superconductors (HTSc) have the highest transition temperatures [1], their practical applications are yet limited. The NbN superconductor with relatively lower (in comparison to HTSc) $T_c$ (~16 K) is being already used for various superconducting applications [2, 3]. Particularly for superconducting devices (tunnel junctions etc.), one needs to employ thin films [4-7]. Niobium Nitride (NbN) thin films have been intensively investigated in the past because of their interesting physical properties and technological applications (superconducting micro-devices, microelectronic catalytic probes and coating) [8]. Crystal structure of the Transition metal nitrides is specified as cubic NaCl (B1-type) with the space group Fm-3m [9]. Particularly NbN have many phases, the cubic fcc $\delta$-NbN, the tetragonal $\gamma$-$Nb_4N_3$, various hexagonal phases $\delta'$-NbN, $\varepsilon$-NbN, and $\beta$-$Nb_2N$ (having low nitrogen content) [10]. In these various nitrides of Nb, only the $\delta$-NbN has been reported as superconducting phase [9-12]. The crystal structure and Nb/N composition affect superconducting properties of NbN. The superconducting NbN has high critical temperature ($T_c$=16 K) with short coherence length ($\xi \approx 5$ nm) and large penetration depth ($\lambda \approx 200$ nm), which allows fabrication of few nano-meter thick superconducting thin films with moderately high $T_c$ [13]. As a conductor in high-field magnet applications sputtered NbN thin films display high-performance superconducting properties. It is possible to improve the properties of NbN thin film through the optimized deposition parameters. Through the optimized deposition conditions one can manipulate the grain size and morphology of the films. This can improve the current density under higher applied magnetic fields [14]. Few hundred nanometer thick films show amorphous or polycrystalline structure [15]. The structural defects, oxidation, and mechanical stress play crucial role for NbN thin films. It is difficult to obtain controlled crystalline structure for the thin films. However, optimization of modern devices can be done from the known superconducting and transport properties of thin NbN films [16]. Along with these properties of NbN superconductor in thin film form has most important role for low temperature electronics, such as for the tunnel junction applications, multilayer films of the type (NbN/AlN/NbN) have been used earlier [17]. For the vacuum microelectronic devices application NbN can be used as cathode [18].

In fact several electronic properties of NbN thin film are favourable for variety of applications. Superconducting transition temperature ($T_c$) of NbN thin films is very sensitive to the growth conditions [19, 20]. NbN has been prepared as thin films using different techniques; including reactive phase DC and RF magnetron sputtering [5, 12, 21], atomic layer deposition (ALD), metal organic chemical vapour deposition (MOCVD) [7] and pulsed laser deposition (PLD) [11,22]. Effect of various deposition fabrication parameters such as argon and Nitrogen pressure in the work chamber, deposition time, choice of substrate and the power supplied to the target on the structural and superconducting properties of the deposited NbN need to be investigated in depth. This work presents results obtained for the growth of NbN thin films on Si (100) substrate, through DC sputtering technique. Also, we studied the effect of Gd doping in superconducting NbN film. The upper critical field $Hc_2(0)$ of the optimized superconducting films is evaluated from the magneto-transport measurements in applied of up to 140 kOe with help of a PPMS (Physical Property Measurement System)

## Experimental

Thin films of NbN and NbGdN were synthesized through reactive dc sputtering (Excel Instruments-INDIA) by sputter Nb metal target of 99.95% purity in an $Ar/N_2$ gas mixture. The substrate temperature and ambient pressure during growth for all the films were fixed at 600 °C and 7mTorr, respectively. Thin film samples were synthesized by keeping deposition time, sputtering power and $Ar/N_2$ gas pressure fixed. The highest $T_c$ (14.8 K) for NbN film is obtained using 250 W sputtering power using Si (100) as substrate [23]. The phase formation of each film is checked through X- ray diffractometer Rigaku (CuKα-radiation). The thicknesses of the films were measured using Stylus profilometer. Morphology and the roughness of the all three films were taken by Atomic Force Microscope (AFM) Multimode V (NS-V) VEECO Instruments Inc. The scanning electron microscopy (SEM) images of the prepared films are taken on a ZEISS-EVO MA-10 scanning electron microscope and Energy Dispersive X-ray spectroscopy (EDAX) is employed for elemental analysis. Resistivity measurements with magnetic field are carried out on the physical property measurement system (PPMS-14T) from Quantum Design-USA.

## Results and discussion

The superconducting NbN and NbGdN films were deposited by *DC* reactive sputtering of a pure Nb and Gd targets in an Ar/N$_2$ gas mixture, at a total pressure of about $10^{-3}$ mbar. The Silicon wafer (100) substrates were kept at 600°C during the film deposition. The thickness *d* of the resulting film was inferred from the sputtering time and a predetermined deposition rate 0.66 nm/s. The film growth was optimized with respect to the partial pressure of N$_2$ and the deposition rate to provide the highest transition temperature for the studied 400-nm thick NbN film.

Figure 1 shows the XRD patterns for NbN and NbGdN thin films for Ar:N$_2$ gas pressure ratios 80:20. From the XRD pattern we observed (111), (200), (220) and (222) plane of $\delta$-NbN also seen is high intensity peak of substrate Si (400). Gd doped NbN film showed only the (111) and (222) planes. All the studied films are free from any oxide impurity. It can be confirmed from the XRD pattern.

Figure 2(a) is shows the SEM image of NbN thin film, indicating the growth of NbN films as a compact granular structure, with an average grain size of 20 nm. From these results, it can be concluded that using sputtering, nano-crystalline NbN films with a crystalline size of approximately 20–50 nm can be prepared. From the EDAX study some oxygen content is seen in the NbN thin film, see figure 2(b). Interestingly the amount of oxide is so small that it is not seen in the XRD pattern of the same. Figure 3(a) shows the SEM image for NbGdN film, indicating the growth of NbN films with Gd, having an average grain size of 20 nm. It is presumed that Gd may act as effective pinning centers and hence relatively higher upper critical field for NbGdN film. Presence of Gd is confirmed from the EDAX study of NbGdN thin film [see figure 3(b)].

Figure 4a and 4b show the topographic AFM images of NbN and NbGdN films. The AFM images consist of irregular submicron-sized features for the sample deposited at 600◦C, and at N$_2$/Ar pressure ratio of 20% to 80%. Very dense and almost uniform in height islands are visible. These islands show an increase in their average height and decrease in their density for NbGdN film. The AFM image revealed better crystalline structure and more uniform size distribution in case of NbN film. The root mean square values (RMS) of surface roughness were obtained from respective AFM images of NbN and NbGdN films with the Nanoscope imaging software. The scanning area was 2×2 $\mu m^2$. NbN film showed a very smooth surface morphology. In case of NbGdN thin film the RMS roughness is increased from 1.10 nm to 2.34 nm. Hence it is concluded that roughness of the NbN film increases with incorporation of the Gd

Figures 5(a) and 5(b) show the Resistivity Vs Temperature plots in varying magnetic fields for NbN and NbGdN films. We observed onset $T_c$ of 14.8K, and 11.2K for NbN and NbGdN film respectively in zero field. The magneto transport measurement up to 14 T applied field for NbN film and 13T for NbGdN film are also shown in Figures 5(a) and 5(b). For both the samples that the transition is sharp in zero field, but with the increasing applied magnetic field the transition width increases slightly. Since the superconducting transition becomes broad in the presence of magnetic field, the experimental values of $T_c$(H) depend on the criterion, one uses to determine $T_c$(H). The upper critical for both samples, have been calculated from the resistive transitions, using the criterion of 90% of the $\rho_n$ value, where $\rho_n$ is the normal resistivity. We have calculated upper critical field by applying the GL theory. From a linear extrapolation of $H_{c2}$(T) to zero temperature one obtains the $H_{c2}$(0). The experimental data for the $H_{c2}$ (T) of NbN thin films can be described with high accuracy by the Ginzburg–Landau (*GL*) equation [24,25]:

$$H_{c2}(T) = H_0 \left[\frac{1-t^2}{1+t^2}\right] \quad \ldots\ldots\ldots\ldots (1)$$

With $t = \frac{T}{T_c}$ and $H_0$ is the thermodynamic critical field at 0°K. Fitted values of upper critical field for both NbN and NbGdN sample are 40Tesla and 31Tesla respectively. These results are plotted in Figure 6. The variation of $H_{c2}$(T) with normalized temperature ($T_c$/T) is shown in inset of figure 6. It is clear from Figure 6 and its inset that Gd doping in NbN superconductor could not improve the upper critical field. Further studies are warranted to control the distribution and size of doped Gd in NbN films so that the same could act as effective pinning centers and enhance the superconducting performance of the host superconductor.

**Conclusion**

We have presented the structural and superconducting properties of NbN and NbGdN thin films deposited on Silicon (100) substrates by DC reactive sputtering technique. Our samples show an excellent crystalline quality of the resulting NbN films with only small quantity of Oxygen embedded in main δ-NbN phase for both samples. *SEM* image for Ar:$N_2$ gas ratio

80:20 showed granular structure, with an average grain size of 20 nm. *AFM* images showed dense and uniform surface of the films with average grain size of 20-30 nm and *RMS* roughness of 1.10 and 2.34 nm for NbN and NbGdN films respectively. We observed onset $T_c$ of 14.8K, and 11.0K for NbN and NbGdN films respectively in the zero field. The upper critical field $Hc_2(0)$ of the superconducting film is evaluated from the magneto-transport measurements to be 40.1 Tesla and 30 Tesla respectively for NbN and NbGdN films. It seems Gd doping did not act as pinning enters in NbN superconducting film.

## Acknowledgements

The authors are grateful for the encouragement and support from Prof. R. C. Budhani (Director NPL). Rajveer Jha would like to thank the CSIR for providing the SRF scholarship to pursue his Ph.D. This work is also financially supported by Department of Science and Technology (DST-SERC) New Delhi, India.

# Figure Captions

**Figure 1** XRD patterns of NbN and NbGdN films deposited on Si (100) substrate.

**Figure 2** (a) SEM image of NbN film shows granular structure. (b) EDAX spectrum for the same film.

**Figure 3** (a) SEM image of NbGdN film shows granular structure. (b) EDAX spectrum for the same film.

**Figure 4** (a) Topographic AFM images of the NbN films deposited at Ar:$N_2$ gas pressures ratios 80:20 (b) Topographic AFM images of the NbGdN films deposited at same condition.

**Figure 5** Superconducting transition zone of resistivity vs temperature plot under applied magnetic field $\rho(T)H$ (a) NbN thin film (b) NbGdN film.

**Figure 6** Upper critical field ($H_{c2}$) versus temperature ($T$) plot for NbN and NbGdN thin film theoretical fitting by the GL equation. Inset shows the same with reduced temperature ($T_c/T$)

**Figure 1**

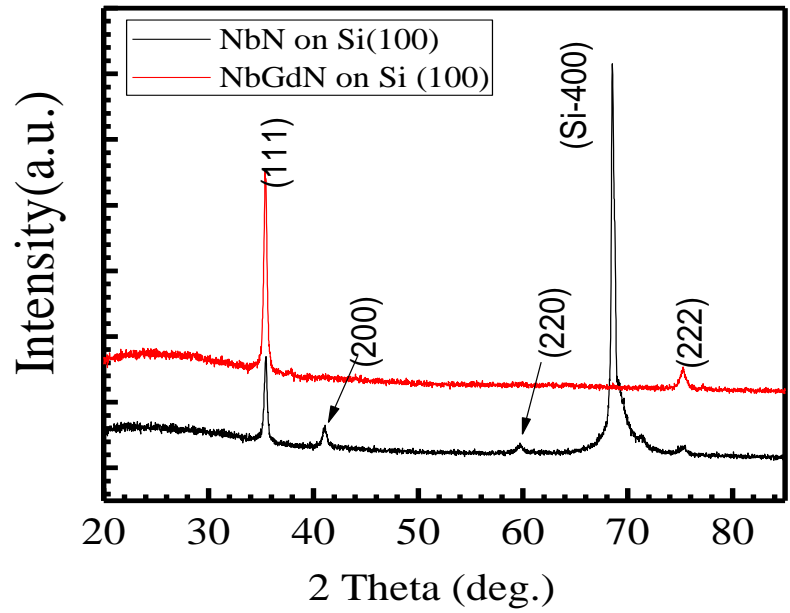

**Figure 2(a)**

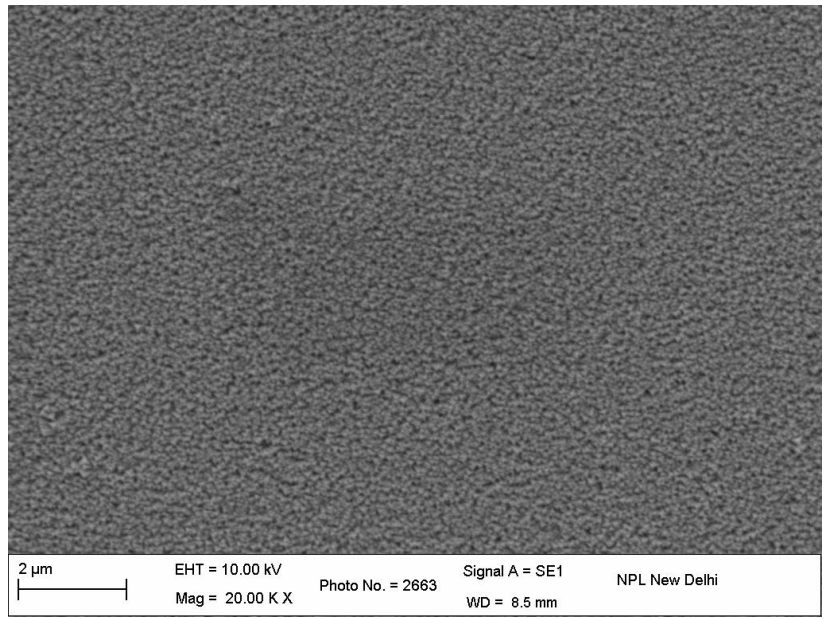

**Figure 2(b)**

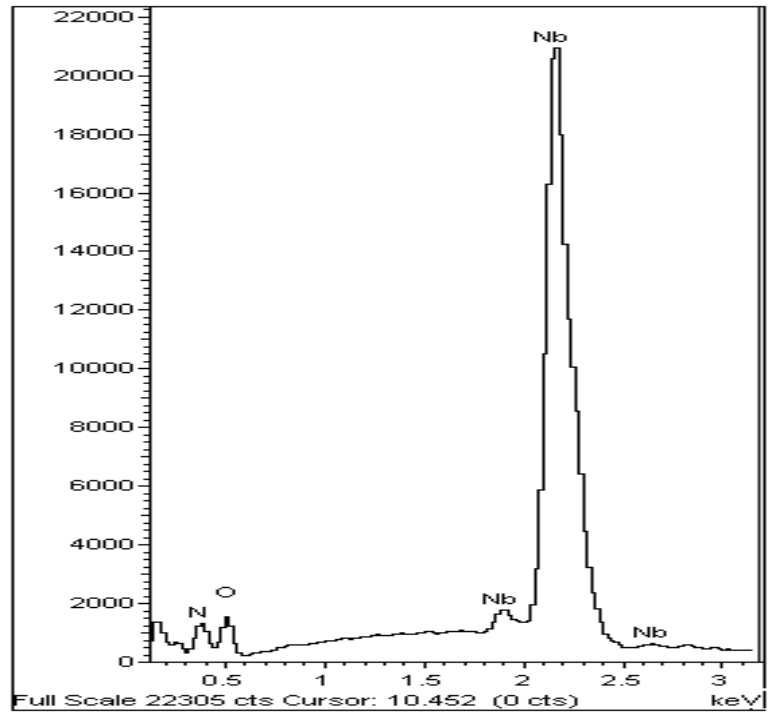

**Figure 3(a)**

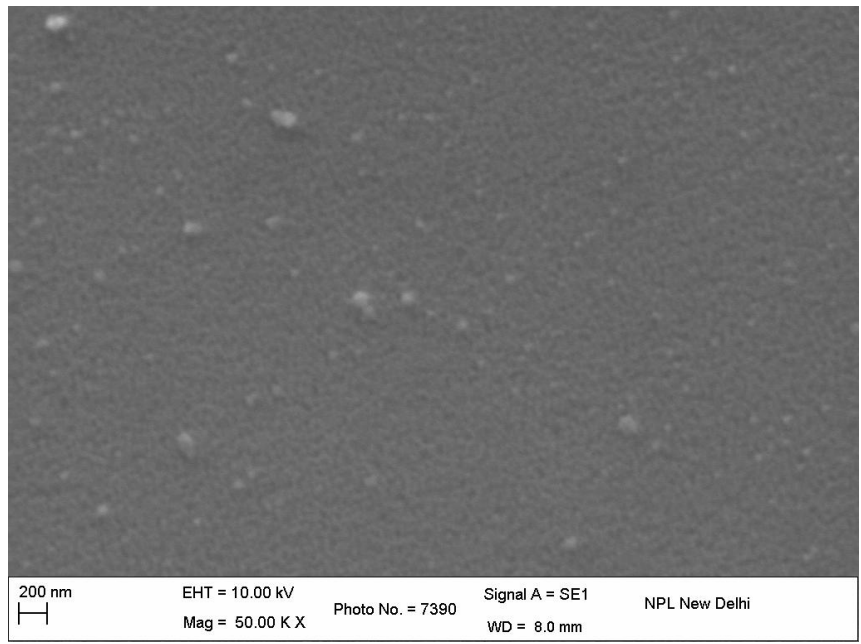

**Figure 3(b)**

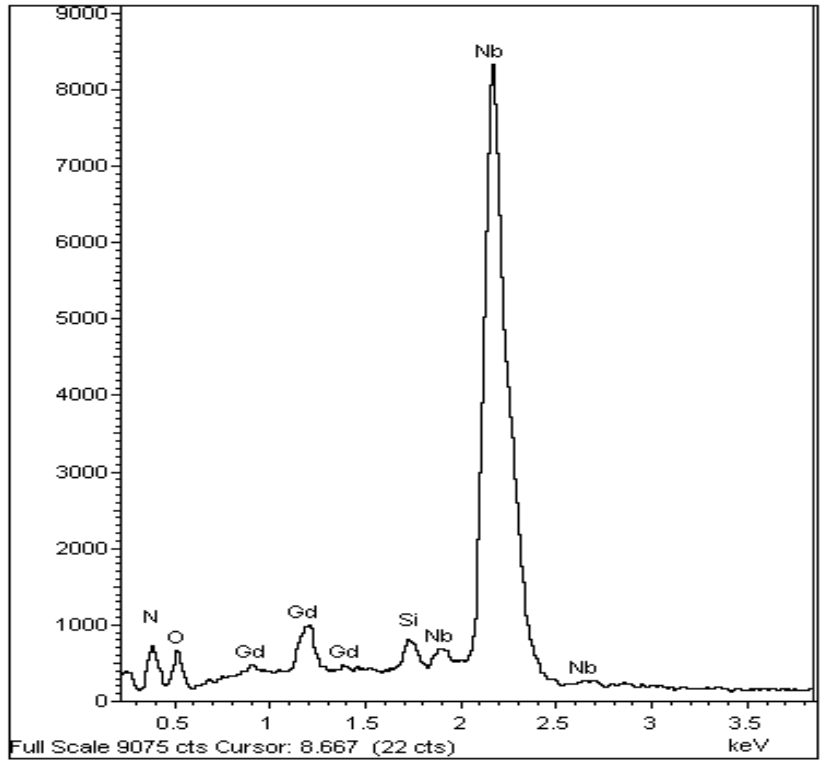

**Figure 4(a)**

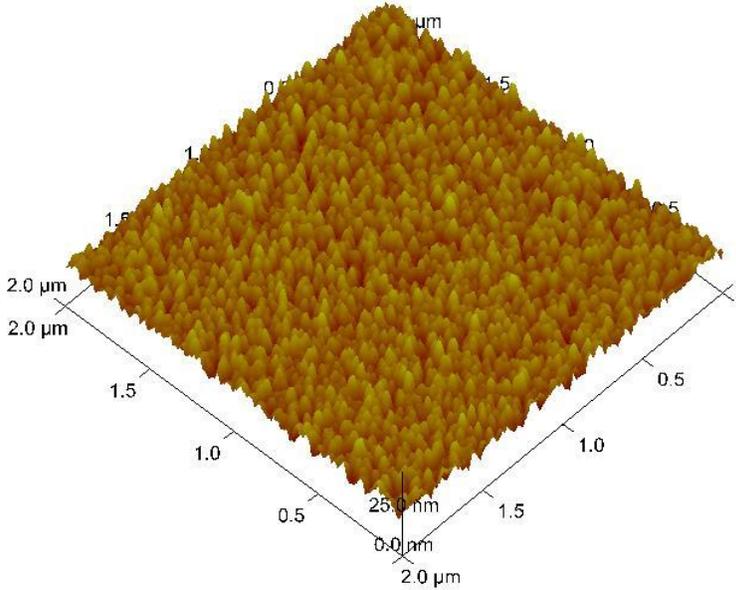

**Figure 4(b)**

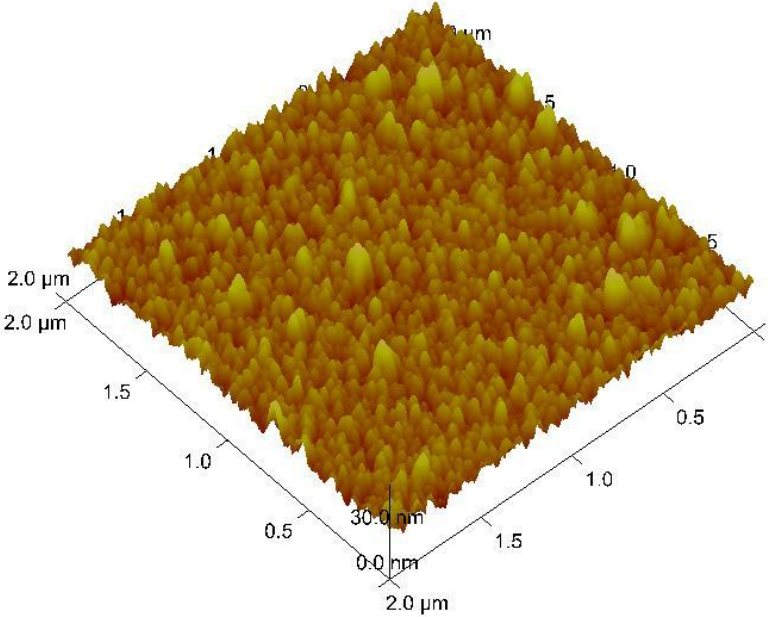

**Figure 5(a)**

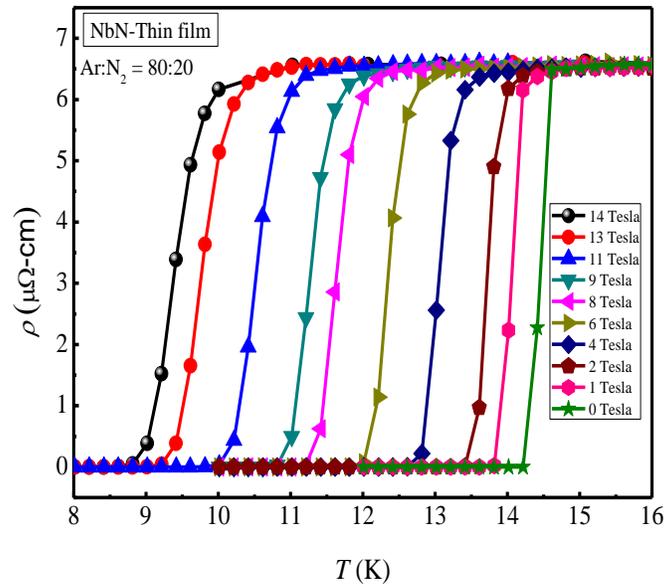

**Figure 5(b)**

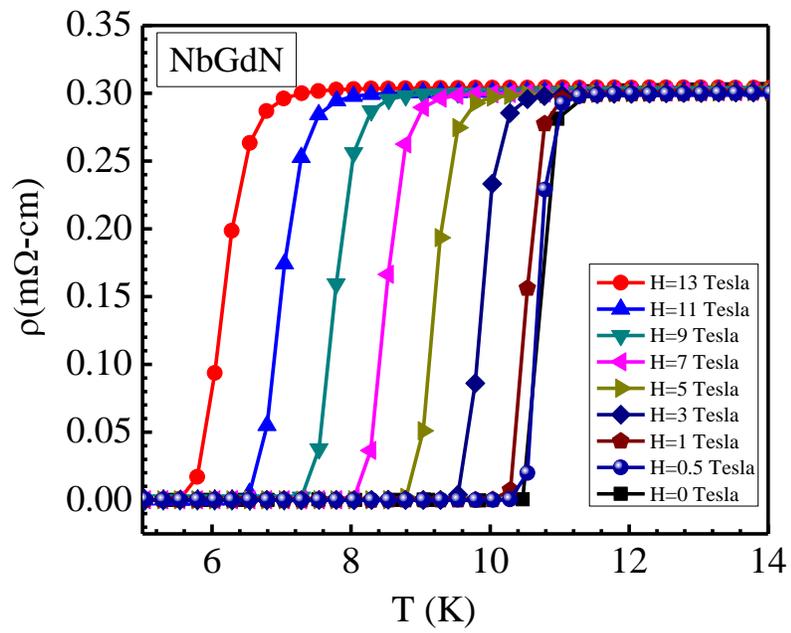

**Figure 6**

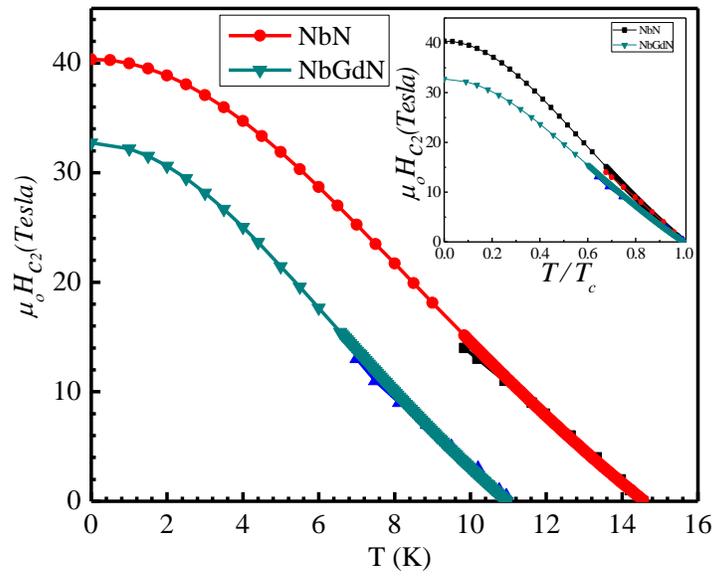